\documentclass[aps,prl,twocolumn,a4paper,10pt]{revtex4-1}

\usepackage[T1]{fontenc}		
\usepackage[english]{babel}		
\usepackage[latin1]{inputenc}	
\usepackage{times}

\usepackage{amsmath}			
\usepackage{amssymb}			
\usepackage{bbm}				

\usepackage[colorlinks=true, a4paper=true, pdfstartview=FitV,linkcolor=blue, citecolor=blue, urlcolor=blue]{hyperref}

\usepackage{graphicx}			
\usepackage{graphics}			
\DeclareGraphicsExtensions{.pdf}
\usepackage{color}

\newcommand{\bea}{\begin{eqnarray}}
\newcommand{\eea}{\end{eqnarray}}

\newcommand{\bk}{\mathbf{k}}

\newcommand{\bq}{\mathbf{q}}
\newcommand{\bsigma}{\boldsymbol{\sigma}}

\begin{document}

\title{Correlations in non-Hermitian systems and Diagram techniques for the steady state}

\author{Johan~Carlstr\"om}
\affiliation{Department of Physics, Stockholm University, 106 91 Stockholm, Sweden}
\date{\today}

\begin{abstract}
We describe a diagrammatic technique for non-Hermitian fermionic systems that is applicable in the steady state, and which allows addressing correlations effects by systematic expansion. 
Applying this method to exceptional points or rings, we find that nodal objects in non-Hermitian systems are generically displaced in momentum-space due to interactions. 
This in turn can be connected to the fact that exceptional points invariably break a class of orthonormal symmetries that are generally present for nodal points in Hermitian systems, and which protect the integrity of the node at low energy scales.  
\end{abstract}
\maketitle


{\it Introduction---} 
Topological semimetals have become the focal point of current research due to a breadth of new forms of quantum matter that includes emergent quasi-particles in the form of Weyl and Dirac fermions that were originally considered in the context of high-energy physics, as well as the realization of their concomitant quantum anomalies \cite{RevModPhys.90.015001, doi:10.1080/00018732.2014.927109}. 
Invoking symmetry arguments, these 
band structures can furthermore be generalized to line-nodes \cite{Bzdusek2016}, knotted band touchings \cite{PhysRevB.96.201305}, higher order Weyl nodes \cite{PhysRevLett.108.266802,Huang1180}, and even Weyl semimetals that exhibit analogues of non-abelian particle statistics \cite{Wueaau8740, 2019arXiv190710611B}. 

Recently, the experimental realization  of topological phases in optical waveguides with dissipation \cite{PhysRevLett.115.040402, Cerjan2019,Weimann2016} has motivated an intense effort to generalize central ideas of topological band-theory to the case of non-Hermitian Hamiltonians \cite{PhysRevLett.121.136802,PhysRevX.8.031079,PhysRevB.99.235112}. 
While a complete understanding of this topic is still lacking, a number of striking differences from the Hermitian case have already been established. Notable examples of these include a modified bulk-boundary correspondence \cite{2018PhRvL.121b6808K,PhysRevLett.121.086803,Xiong_2018}, reduced co-dimension of band touching points \cite{2004CzJPh..54.1039B}, bulk Fermi arcs in 2D \cite{Zhou1009} and Fermi ribbons in 3D \cite{PhysRevA.98.042114,PhysRevB.99.161115}, which are direct manifestations of non-abelian statistics in the non-Hermitian regime \cite{Heiss_2012}. 

Technological applications based on non-Hermitian band topology are currently focused on sensors and detectors that exploit the sub-linear dispersion in the proximity of exceptional points to enhance signal response.
Presently, implementations utilizing optical micro-cavities already exist
  \cite{Chen2017}, while proposals based on ultra-cold atomic gases were put forth recently \cite{PhysRevA.99.011601,PhysRevA.99.063616}. 
  In the latter implementation, a central idea is to exploit particle correlations to realize higher order exceptional points with even steeper susceptibilities to a weak signal. 
While the particular models considered in \cite{PhysRevA.99.011601,PhysRevA.99.063616} are exactly solvable, the extension of non-Hermitian physics to cold atomic gases and electronic systems out of equilibrium, where interactions are ubiquitous \cite{PhysRevLett.118.045701,Li2019,PhysRevA.99.031601,PhysRevA.94.053615,Tomitae1701513,PhysRevResearch.1.012003}, represents a fundamental theoretical challenge since the current frameworks of quantum many-body theory either rely on unitary time-evolution, or are constructed at thermal equilibrium.
Note that these works should not be confused with a number of recent reports on exceptional points in electronic systems at equilibrium that result from a complex self energy \cite{PhysRevB.97.041203,PhysRevB.99.041116,PhysRevB.98.035141}. 

In this work we explore correlation effects in non-Hermitian systems, in particular demonstrating how the steady state can be described using diagrammatic techniques. 
Applying this method to simple nodal objects such as an exceptional point or ring, we find that these are generally more susceptible to interactions than band touching points in Hermitian systems. This in turn is related to the fact that exceptional points invariably break a class of orthonormal symmetries that are generally present for conventional nodal points.

{\it Perturbation theory---} 
In Hermitian systems, the perturbative expansion is generally organized in a diagrammatic series, which provides a systematic way of computing corrections to observables either in the ground state, or at thermal equilibrium. 
While dissipative or driven systems are by construction neither in the ground state or at equilibrium, it is still possible to conduct 
a perturbative expansion in the large-time limit, provided the existence of either a single steady state or a set of such states that are {\it macroscopically indistinguishable}. 

More precisely, this condition may be understood as follows: 
Degeneracies of the steady state implies the existence of zero-decay modes for which the imaginary part of the energy vanishes exactly, that thus neither decays or grows.
If the i-Fermi surface, which separates growing and decaying modes, has a non-zero co-dimension with respect to momentum-space, then the difference in occupancy between two steady states is non-extensive in the system size, and can therefore not be registered by macroscopic observables. 
By contrast, if the i-Fermi surface has the same dimension as the system, then the difference in occupancy between steady states becomes extensive, which violates the condition. 
The latter scenario which can arise in systems that posses certain spectral symmetries \cite{PhysRevB.99.041406,PhysRevB.99.041202} or due to fine-tuning must be treated separately and will not be considered hereafter. 

To derive a description of the steady state, we start by assuming an initial state of our system at $t=0$ that is denoted by
\bea
|\psi_0\rangle. \label{initial state}
\eea
The time evolution of this state is then described by
\bea
\langle\psi_i(t)|=\langle\psi_i|e^{itH^\dagger},\;|\psi_i(t)\rangle=e^{-itH}|\psi_i\rangle,\;t\ge 0  \label{Ut}
\eea
giving an expectation value of the operator $\hat{O}$ according to
\bea
\langle \hat{O(t)}\rangle =\frac{\langle \psi_0|  e^{itH^\dagger} \hat{O}e^{-itH} |\psi_0\rangle }{\langle  \psi_0| e^{itH^\dagger} e^{-itH}|\psi_0\rangle}. \label{Ot}
\eea 

 Since a non-Hermitian Hamiltonian generally lacks an orthonormal eigenbasis, it follows that the time-evolution operator is typically not diagonalizable. 
If there are no exceptional points in the theory then it can still be expressed in terms of biorthogonal eigenvectors, though there is also a scenario when the eigenbasis of $H$ does not even span the entire Hilbert space. 
When constructing the time-evolution operator, we must therefore distinguish between states that are within or outside the space of eigenvectors:
\bea
 H |\alpha_i,R\rangle = \lambda_i |\alpha_i,R\rangle, \;
H|\beta_i,R\rangle = \lambda_{ij}|\psi_j,R\rangle,
\eea
where $\{|\alpha_i,R\rangle\}$ are the right eigenvectors of $H$, $\{|\beta_i,R\rangle\}$ spans the reminder of the Hilbert space, and $\{|\psi_i,R\rangle\}=\{|\alpha_i,R\rangle\}\cup \{|\beta_i,R\rangle\}$ forms a complete basis. 

Let us now return to the initial assumption that the system is described by a set of steady states which contains elements that are more rapidly growing (or slowly decaying) than the remaining states, and which are sufficiently similar that they may not be discriminated between by macroscopic observables. 
 In the large-time limit we may drop sub-leading terms, which gives a time-evolution operator of the form
\bea\nonumber
\lim_{t \to\infty} U(t) = \sum_j e^{-i t \epsilon_j}|\alpha_j,R\rangle\langle  \alpha_j,L|,  		\\
\lim_{t \to\infty} U^\dagger( t) =\sum_j e^{it\epsilon_j^*}|\alpha_j,L\rangle\langle  \alpha_j,R|,		\label{evol}
\eea
which projects onto a subspace where $\Im (\epsilon_j)=\Im (\epsilon_k)$. 
Since the states in (\ref{evol}) are macroscopically indistinguishable, we may in principle give preference to a specific state by lifting the degeneracy in (\ref{evol}) without affecting macroscopic observables. This can be achieved by adding a convergence factor of $\pm i\eta$ to the energy of the modes situated exactly on the i-Fermi surface. 
After taking $t\to \infty$ we proceed to take $\eta\to 0$, so that we project on a single state which has definite particle occupation number.   

Returning to the expectation value (\ref{Ot}), we now recognize that we can write the particle density in the steady state according to 
\bea
\lim_{t\to\infty} \langle \hat{n}(t,\bk)\rangle=\frac{\langle \alpha_0,R|\hat{n}(\bk)|\alpha_0,R\rangle}{\langle \alpha_0,R|\alpha_0,R\rangle}, \label{Nt}
\eea
where $|\alpha_0,R\rangle$ is the sole steady state once we have lifted the degeneracy in (\ref{evol}). Since this state has definite particle occupation number, it follows that it is an eigenstate of the number operator $\hat{n}(\bk)$. Using (\ref{evol}) we can thus write (\ref{Nt}) on the form
\bea
\lim_{t\to\infty} \langle \hat{n}(t,\bk)\rangle=\frac{\langle \alpha_0,L|\hat{n}(\bk)|\alpha_0,R\rangle}{\langle \alpha_0,L|\alpha_0,R\rangle}\\
=\lim_{t\to\infty}  \frac{\langle \psi_0| e^{-it H}\hat{n}(\bk)e^{-itH}|\psi_0\rangle}{\langle \psi_0| e^{-itH}e^{-itH}|\psi_0\rangle}, \label{Ntfinal} 
\eea
where we have assumed that $\psi_0$ is not orthogonal to the steady states, a scenario which in principe would require retaining subleading terms of (\ref{evol}).
Expressing $\hat{n}$ in terms of field operators and conducting a time translation, we obtain two-point correlators on the form
\bea
\langle \Psi^\dagger(t_2)\Psi(t_1)\rangle
=\lim_{t\to\infty} \times
\\
\frac{\langle \!\Psi_0|e^{-i(t-t_2)H} \Psi^\dagger e^{-i(t_2-t_1)H} \Psi e^{-(t+t_1)}|\psi_0\rangle}{\langle \Psi_0|e^{2t}|\psi_0\rangle}.\label{twoPoint}
\eea
At this point we will make a second assumption, namely that the Hamiltonian can be decomposed into a bilinear part $H_0$ and an interaction part $H_1$,
\bea
H=H_0+H_1, \label{bilinear}
\eea
which in principle allows us to make contact with conventional zero temperature diagrammatics \cite{fetter}. Expanding (\ref{twoPoint}) in $H_1$ we obtain the following series expansion for the full Greens function:
\bea\nonumber
iG(x_1,x_0) =\frac{1}{Z}\sum_n \frac{(-i)^n}{n!}\int_{-\infty}^{\infty} dt_1...dt_n \;\;\;\;\\
\!\times\!\langle \psi_0|e^{\!-\!it\! H_0} T_t [H_1(\!t_1\!)\!...H_1(\!t_n\!)\Psi(\!x_1\!) \Psi^\dagger(\!x_0\!)]e^{\!-\!it\! H_0}|\psi_0\rangle, \label{iG}\;\;\;\;
\eea
with $t\to\infty$, which can be expressed in terms of time-dependent field operators of the form
\bea
\Psi(t)=e^{i t H_0} \Psi e^{-i t H_0}.\label{fieldOperators}
\eea
While the operators (\ref{fieldOperators}) are described by non-unitary time-evolution, bilinearity of $H_0$ implies that this is only reflected in the presence of complex exponents, so that for example
\bea
\Psi^\dagger(\bk,t)=e^{i t H_0} \Psi^\dagger(\bk)  e^{-i t H_0} =\Psi^\dagger(\bk) e^{it \epsilon_\bk},\; \epsilon_\bk\in \mathbb{C}\;\;\;\;
\eea
where we recall that the energy $\epsilon_\bk$ is by contrast real in Hermitian systems. Correspondingly, under the assumption that $H_0$ is bilinear, we can treat (\ref{iG}) with Wicks theorem to produce an expansion in terms of connected diagrams that relates the full Greens function to its bare counterpart, which is given by
\bea
\!iG_0\!(t_1\!-\!t_2,\!\bk)\!=\!\frac{\!\langle \psi_0\!|e^{\!-\!itH_0\!} T_t \Psi(\bk\!,t_1\!) \Psi^\dagger\!(\bk\!,t_2\!)e^{\!-\!itH_0}\!|\psi_0\!\rangle\!}{\langle \psi_0|e^{-i2tH_0}|\psi_0\rangle},\label{iG0}\;\;\;\;\;\;
\eea
where $t\to\infty$. Notably, (\ref{iG0}) is independent of the initial state as long as it has a nonzero overlap with the steady state of $H_0$. Under this assumption, the state which is acted upon by the field operators is either vacant or occupied, depending on whether it is characterized by gain or loss at the level of bilinear theory. 
 Specifically, we obtain
\bea\nonumber
iG_0(t,\bk) =-\theta\big(\Im[\epsilon(\bk)]\big)e^{-it \epsilon(\bk)}, \;t<0\\
iG_0(t,\bk) =  \theta\big( -\Im[\epsilon(\bk)]\big)e^{-it \epsilon(\bk)},  \;t>0 
\label{G0struc}
\eea 
where $\Im (\epsilon)=0$, at the level of bilinear theory, defines the i-Fermi surface that separates states with gain and loss.
Since the energy is complex, the Greens function is exponentially localized in time, and so the time integral convergences naturally without a convergence factor $\sim  i\eta$. 
 In this respect, the expansion is reminiscent of Matsubara formalism in the zero temperature limit. 
Taking the Fourier transform of (\ref{G0struc}), we obtain the bare frequency-Greens function,
\bea
G_0(\omega,\bk)=\frac{1}{\omega-H_0(\bk)}.\label{Gomega}
\eea
While superficially similar to the zero-temperature Greens function of a Hermitian system, (\ref{Gomega}) generally exhibits poles at a finite distance from the real axis that reflect the dissipative nature of non-Hermitian systems. 
  In the limit $t\to -0$, the full Greens function (\ref{iG}) reproduces the particle density of the steady state in accordance with (\ref{Ntfinal}).
  
{\it Fock theory---}
Having established a diagrammatic framework, we can now return to the prelusive question of how correlations affect nodal points in the non-Hermitian regime. At the lowest order, we obtain two types of diagrams, namely Hartree and Fock type corrections. Of these, the former typically only lead to a renormalization of the chemical potential, while the Fock term gives a nontrivial correction to the self energy of the form 
 \bea\nonumber
 \Sigma_{\text{Fock}}(\bk)=\int d\bq  d\omega V(\bq) i G^0(\omega,\bk-\bq)\;\;\;\;\\
 \!=\!\int\! d\bq  d\omega V\!(\bq) \! \frac{i}{\omega \!-\!H_0(\bk\!-\!\bq)}\!
  =\!\int\! d\bq  d\omega V\!(\bq) i\! \frac{\omega\!+\!H_0}{\omega^2\!-\!H_0^2},\;\;\;\;
 \eea
where we have absorbed a factor $(2\pi)^{-D}$ into $V$.
Dropping the part which is odd in frequency and assuming a particle-hole symmetric two-band model we obtain  
\bea
i \int d\omega \frac{H_0}{\omega^2-H_0^2}=i \int d\omega \frac{H_0}{(\omega-\Delta)(\omega+\Delta)},
 \eea
 where we have defined $\Delta^2\sigma_0=H_0^2$. If we choose to define $\Delta$ such that $\Im (\Delta)>0$, then we obtain a pole in the upper half-plane corresponding to
$\omega=\Delta$. This gives a Fock integrand on the form
\bea
\xi(\bk)=i H_0(\bk)\frac{i 2\pi}{2 \Delta(\bk)} =- \pi\frac{H_0(\bk)}{\Delta(\bk)},\label{Xi}
\eea
with
\bea
\Sigma(\bk)= \int d\bq V(\bq) \xi(\bk-\bq).\label{defXi}
\eea 
Thus, we see that the Fock integrand, despite being non-Hermitian, possesses a real spectrum. It should be stressed however that it is still generally accompanied by a non-orthogonal eigenbasis which it shares with $H_0$.

To examine the correction from (\ref{Xi}) to an exceptional point, we consider a minimalistic model of the form
\bea
H_0=\kappa+\gamma,\;
\kappa=\sigma_x \!+\! i\sigma_z,\;\gamma= \frac{k_x}{2}\sigma_x\!+\!\frac{k_z}{2}\sigma_z 
\eea
which possesses an exceptional point in $k_x=k_z=0$, where $\gamma$ vanishes. 
To leading order, this gives a gap which can be expressed on polar form according to
\bea
\Delta=\sqrt{k_x+i k_z}= \chi e^{i\phi/2},\; \chi=|k_x+i k_z|^{1/2}=\sqrt{k}.\;\;\;\;
\eea
For a contact interaction we have $V(\bq)=V$, implying that (\ref{defXi}) corresponds to an integral of the Fock integrand $\xi(\bq)$ over $\bq$. In this scenario we can work in polar coordinates, and symmetrize the integrand with respect to the angle $\phi$. Decomposing $\xi$ into terms $\sim \kappa$ and $\sim \gamma$, we obtain 
\bea
\xi_\kappa^s=\int_{0}^{2\pi} kd\phi (-\pi \kappa \frac{e^{-i\phi/2}}{\chi})=4i\pi\kappa\sqrt{k},\;\;\;\;\\
\xi_\gamma^s\!=\!\int_0^{2\pi} \!kd\phi (-\pi  \gamma \frac{e^{-i\phi/2}}{\chi})\!=-\frac{2\pi k^{3/2}}{3 }(i\sigma_x\!+\!2\sigma_z).\;\;\;\;
\eea
In the next stage we integrate over $k=\sqrt{k_x^2+k_z^2}$ to obtain
\bea
J_\kappa = V\int dk \; 4i\pi  \sqrt{k},\; J_\gamma = -V\int dk \; \frac{2\pi k^{3/2} }{3},\label{J}
\eea
which gives a renormalized dispersion of the form
\bea
H_0+\Sigma= (1+J_\kappa)\kappa+ \sigma_x\frac{k_x+2iJ_\gamma}{2}+ \sigma_z\frac{k_z+4J_\gamma }{2}.\;\;\;\;\label{H0effective}
\eea
Computing the spectrum of (\ref{H0effective}) we find a node in
\bea
k_x=0, \;k_z=-6J_\gamma,
\eea
so that the exceptional point is translated in momentum space along the $k_z$-axis by a distance that is linear in the interaction strength $V$. 

In principle the finding that correlation effects lead to a translation of nodes in momentum space also holds for an exceptional ring that results when adding a non-Hermitian perturbation to a Weyl point. To the lowest order we then obtain a dispersion of the form
 \bea
 H_0=\bk\cdot \bsigma+ i\sigma_z, 
 \eea
with a gap that is given by
\bea
\Delta=\frac{k_z}{|k_z|}\sqrt{\bk^2-1+2ik_z}.
\eea
Similarly to in the preceding scenario, the integral of $\xi$ over $\bq$ is simplified by symmetrization of the integrand. To this end we thus note the following properties of the dispersion and the gap 
\bea
\frac{H_0(\bk_{xy},k_z)+H_0(-\bk_{xy},k_z)}{2}=(k_z+i)\sigma_z\label{piRot}\\
\Delta(\bk_{xy},k_z)=\Delta(-\bk_{xy},k_z)\label{DpiRot}\\
\Delta^*(\bk_{xy},k_z)=-\Delta(\pm\bk_{xy},-k_z)\label{reflection}
\eea
Using the properties (\ref{piRot},\ref{DpiRot},\ref{reflection}) we can construct a Fock integrand that is symmetrized with respect to a $\pi$-rotation around the $z-$axis according to
\bea
\xi_{\pi}=\frac{\xi(\bk_{xy},k_z)+\xi(-\bk_{xy},k_z)}{2}=-\frac{(k_z+i)\sigma_z}{\Delta(\pm \bk_{xy},k_z)},\label{Xipi}
\eea
which is thus diagonal, but features a complex spectrum. In the next stage, we use (\ref{reflection}) and symmetrize with respect to a reflection in the $xy-$plane. We begin by noting the relationship
\bea
\xi_{\pi}(\bk_{xy},-k_z)=-\frac{(-k_z+i)\sigma_z}{-\Delta^*(\bk_{xy},k_z)}=\xi_{\pi }^\dagger(\bk_{xy},k_z)
\eea
that allows us to define the rotation and reflection symmetrized Fock integrand according to
\bea
\xi_s(\bk)\!=\!\frac{\xi_{\pi }(\bk_{xy},k_z)\!+\!\xi_{\pi }(\bk_{xy},-k_z)}{2}\!=\!\frac{\xi_{\pi }(\bk)\!+\!\xi_{\pi }^\dagger(\bk)}{2}.\;\;\;\; \label{Xis}
\eea
The implication of (\ref{Xipi}) and (\ref{Xis}) is that $\xi_s(\bk)$ is a diagonal matrix with real eigenvalues of the form
\bea
\xi_s(\bk)=f(\bk)\sigma_z,\;f\in \mathbb{R},\;f(\bk)<0 \;\forall \bk\label{XisRing}
\eea 
Taking $V(\bq)$ to be a contact interaction, inserting (\ref{XisRing}) in (\ref{defXi}) and integrating over $\bk$, we find that the exceptional ring is translated along the $k_z-$axis by 
\bea
\Delta k_z=-V\int d\bk f(\bk),
\eea
which is orthogonal to the ring.

{\it Orthonormal symmetry breaking---}
The nontrivial response of exceptional points to many-body effects is strikingly different from conventional nodal points, which are at the lowest energy scales generally protected by an orthonormal symmetry. 
Specifically, expanding the dispersion of a Hermitian system to the lowest order around a band touching point, 
we typically find an effective description which is odd under some orthonormal map according to
\bea
H_0(T_-\bk)=-H_0(\bk),\; |T_- \bk|=|\bk|.\label{TM}
\eea
Notable examples of this include Weyl points, where the dispersion is odd under inversion,
\bea
H_0(\bk)=k_i v_{ij}\sigma_j,\; H_0(\bk)=-H_0(-\bk),
\eea
and higher order Weyl semimetals of the form
\bea
H_0(\bk)=(k_x+ik_y)^n \sigma_+ +(k_x-ik_y)^n \sigma_- +k_z\sigma_z
\eea
where the orthonormal map corresponds to rotation around the $z-$axis and reflection in the $xy-$plane. Further examples arise in single and multilayer graphene. 

For two-body interactions, the property (\ref{TM}) implies a symmetry of the self energy given by
\bea
\Sigma(\omega,\bk=0)=-\Sigma(-\omega,\bk=0),
\eea
which in turn results in a solution to Dysons equation at zero energy \cite{PhysRevB.97.161102}.
Thus, not only is the semimetallic phase protected, but the position of the node in momentum-space is preserved by the symmetry (\ref{TM}). This fact has also been verified by diagrammatic Monte Carlo simulations \cite{PhysRevB.98.241102}.
Once higher order terms in the dispersion are included, the orthonormal symmetry is generally broken, but the fact that (\ref{TM}) holds to the lowest order implies that interaction effects are significantly diminished in Hermitian nodal systems. 

In contrast to nodal points in Hermitian systems, exceptional points are irreconcilable with the orthonormal symmetry (\ref{TM}), since a vanishing (or diagonal) $H_0$ necessarily possesses an eigenbasis that spans the Hilbert space. 
The implication of this discrepancy is that exceptional points already at the level of linearized theory are shifted in momentum-space due to many-body corrections. In full lattice models this suggests that non-Hermitian systems are far more susceptible to interactions than their Hermitian counterparts, particularly if the interaction is long ranged in real-space and thus rapidly decaying in momentum space.

{\it Discussion---} 
In this work, we have examined the role of correlation effects in non-Hermitian systems, in particular showing how diagrammatic techniques can be used to describe the steady state. 
Although the corrections that transpire from this treatment are formally similar to those of conventional zero-temperature formalism, this class of theories feature a bare Greens-function that generally possesses poles situated at a finite distance from the real axis, implying that it becomes exponentially localized in the time-domain.

The employment of diagrammatic methods in the non-Hermitian regime does not only pave the way for conventional perturbative treatment, but also implies that these systems can be addressed with diagrammatic simulation techniques \cite{VANHOUCKE201095} 
that in principle allow many-body effects to be computed systematically if the series is convergent, even for strongly interacting quantum matter \cite{PhysRevB.97.075119}.

The presence of complex particle energies generally implies that in the limit of large $t$, the time-evolution operator takes the form of a projection onto a subspace, effectively destroying some of the information about the initial state. 
In this context, the prelusive assumption that any two steady states cannot be discriminated between should be understood from the fact that all information about the initial state that can be extracted by macroscopic observables is lost, and this also explains why the expansion is independent of the initial state as long as it has a non-zero overlap with a steady state. 
Correspondingly, if the assumption is violated and the i-Fermi surface has the same dimension as momentum-space, then the time-evolution operator fails to erase measurable thermodynamic properties that are encoded in the initial state. 
In this scenario, additional selection rules based on the real part of the spectrum or a specific choice of the initial state are necessary to compute relevant observables. 

Finally, we note that displacement of nodal points are ubiquitous in interacting non-Hermitian systems except in the case of very particular inter-particle forces, notably when the potential is completely local in momentum space. This fact, which is in turn related to the breaking of an orthonormal symmetry raises the question of whether these corrections can be connected systematically to topological characteristics of the nodal objects \cite{PhysRevLett.123.066405,PhysRevB.100.075403}. 

{\it Acknowledgments---}
This work was supported by the the Swedish Research Council (VR). 
The author would like to thank Emil~J. Bergholtz for important input and discussions.

\bibliography{biblio}

\end{document}